\documentclass{article}
\usepackage{graphicx} 
\usepackage{xcolor}
\usepackage{enumitem}
\usepackage{amsmath,amssymb,tensor}

\usepackage{hyperref}
\hypersetup{
	colorlinks=true,
	linkcolor=blue,
	filecolor=magenta,      
	urlcolor=cyan,
	citecolor=black!30!green,
	pdftitle={Overleaf Example},
	pdfpagemode=FullScreen,
}

\newtheorem{example}{Example}
\newtheorem{remark}{Remark}
\newtheorem{question}{Question}
\newtheorem{theorem}{Theorem}
\newtheorem{definition}{Definition}

\title{On Haantjes tensors for second-order superintegrable systems}
\usepackage{authblk}
\author[1]{Ian Marquette}
\author[2]{Damien McLeod}
\author[3]{Serena Scapucci}
\author[4]{Andreas Vollmer}
\affil[1] {\small La Trobe Department of Mathematical and Physical Sciences, La Trobe University, Bendigo, VIC 3552, Australia.
	E-mail: \texttt{i.marquette@latrobe.edu.au}}
\affil[2] {\small  School of Mathematics and Statistics, University of Sydney, Camperdown, Sydney, NSW 2006, Australia.
	E-mail: \texttt{dmcleod@maths.usyd.edu.au}}
\affil[3]{\small Friedrich-Schiller-Universität Jena, Ernst-Abbe-Platz 2, 07743 Jena, Germany.
	E-mail: \texttt{serena.scapucci@uni-jena.de}}
\affil[4]{\small Universität Hamburg, Fachbereich Mathematik, Bundesstr. 55, 20146 Hamburg, Germany. E-mail: \texttt{andreas.vollmer@uni-hamburg.de}}
\date{9 December 2024}

\begin{document}
	
	\maketitle

	\begin{abstract}
		The vanishing of the Haantjes tensor is an important property that has been linked, for instance, to the existence of separation coordinates and the integrability of systems of hydrodynamic type.
		We discuss the vanishing of the Haantjes tensor for operator fields that admit a large number of so-called conservation laws. In particular, we investigate Haantjes-zero Killing tensor fields that are associated with second-order superintegrable systems.
	\end{abstract}
	
	\section{Introduction}
	
	Let $(M,g)$ be a (pseudo-)Riemannian manifold. 
	Loosely speaking, a superintegrable system on $M$ is defined by a Hamiltonian $H:T^*M\to \mathbb R$ together with a large number of functionally independent functions $F:T^*M\to\mathbb R$ (``constants of the motion'' or ``(first) integrals'') that are constant along solution trajectories of Hamilton's classical equations of motion. Usually, additional assumptions are imposed onto these integrals. For example, we may assume that they are quadratic polynomials in the canonical momenta coordinates (i.e.\ fibre coordinates of canonical Darboux coordinates on $T^*M$).
	The study of superintegrable systems has a long history. It can be traced back to early works, such as the classical Kepler problem in classical mechanics and the Laplace-Runge-Lenz vector \cite{Hamilton,Laplace,Runge,Lenz,Goldstein1975,Goldstein1976}, and to the hydrogen atom in the context of quantum mechanics \cite{FMSUW1966,WSUF1966}. St\"ackel equivalence and coupling constant metamorphosis are important transformations of superintegrable systems, which can be traced back to classical works by Maupertuis and Jacobi \cite{Jacobi,Maupertuis,Blaszak2012,Blaszak2017,Tsiganov}.
	
	Beginning with investigations on the $2$-dimensional Euclidean space in the 1960s \cite{FMSUW1966,FSUW1967,MSVW1967}, the classification of classical and quantum superintegrable systems of second order on surfaces was completed around 2006 \cite{KKPM,KKM-I}.
	For second order maximally superintegrable  systems, altogether 59 superintegrable systems on different spaces  of constant and non constant curvature were identified \cite{KKM-I}, falling into twelve equivalence classes under St\"ackel equivalence \cite{KKM-II}.
	These Hamiltonians have since been linked, for example, to hypergeometric orthogonal polynomials organized in the Askey-Wilson scheme \cite{KKM-askey}, and to Wigner-\.{I}n\"on\"u contractions \cite{KM2014}.
	For a review of many of these developments and an extensive list of references, we refer the reader to \cite{KKM-book,post_2013}.
	
	Second order maximally superintegrable systems possess many interesting properties in the context of classical and quantum mechanics, such as multi-separability of the Schr\"odinger and Hamilton-Jacobi equations \cite{FMSUW1966,WSUF1966}, or exact solvability (both analytic and algebraic) \cite{TTW2001}.
	In the context of classical mechanics, the bounded trajectories are closed, and the motion is periodic \cite{N1972,MW2007}.
	In the quantum case, the underlying symmetry algebra takes the form of a so-called quadratic algebra, allowing one to obtain their degenerate spectrum \cite{dask2001}.
	Recently, algebraic solvability from these quadratic algebras was completed \cite{mar2023}.
	The classification has been extended in various directions, such as for Hamiltonians with magnetic field, involving spin interaction, or possessing higher order integrals \cite{G2004,GW2002,MW2007}.
	
	Restricting again to the case of second order superintegrability, a classification on $3$-dimensional space is to date only partly achieved. A classification for non-degenerate systems has been completed \cite{Capel}. For degenerate and semi-degenerate systems some partial results are known \cite{Evans,ERM17,KKPM}. Some families in $n$-dimensional spaces are also known.
	Generally speaking, the classification problem is rather challenging, as the equations involved become increasingly cumbersome to manage with increasing dimension. In recent years, however, advances have been made thanks to novel geometric techniques \cite{KS19,KSV23,KSV24}.
	These provide a pathway towards further insight into the properties of these systems.
	
	In parallel to this development, developments have been made in the context of integrable hydrodynamical systems, employing Nijenhuis and Haantjes Killing tensors, for instance \cite{FM}. Such techniques have also lead to a better understanding of separation systems \cite{KKM}.
	In spite of this recent progress, many aspects are still unexplored and deserving of further research. To help fill this gap in the literature, and to bring various of these novel geometrical approaches together is one of the purposes of this paper. 
	We begin by recalling some definitions \cite{KKM-book,post_2013}.
	\begin{definition}
		A system with Hamiltonian $H$ is integrable if it admits $n$ constants of the motion $F_1=H$, $F_2$,..., $F_n$ that are in involution $\{F_j,F_k\}=0$, $1 \leq j,k \leq n$ and are functionally independent. Here $\{\, \cdot \, ,\cdot\,\}$ is the Poisson bracket. A Hamiltonian systems is (polynomially) integrable if it is integrable and the constants of the motion are each polynomials in the momenta globally well defined except possibly for singularities on lower dimensional manifolds.
	\end{definition}

	\begin{definition}
		A classical Hamiltonian system in dimension $n$ is (polynomially) superintegrable, if it admits $n+k$ (with $k=1,...,n-1$) functionally independent constants of motion that are polynomial in the momenta and are globally defined except possibly for singularities on a lower manifold. It is minimally (polynomially) superintegrable if $k=1$ and maximally (polynomially) superintegrable if $k=n-1$.
	\end{definition}
	
	\begin{remark}
		The case of dimension two is special and all superintegrable systems in dimension two are maximally superintegrable. Superintegrable systems admitting $2n-1$ functionally independent second order integrals of the motion (i.e.\ taking the form of second order polynomials in momenta) are called second order or quadratically superintegrable systems.
	\end{remark}

	Let $(M,g)$ be a (pseudo-)Riemannian manifold. We refer to a $(1,1)$-tensor field $A$ on $M$ as \emph{operator field}.
	A symmetric $(0,2)$-tensor field $K$ on $M$ with the property that
	$$ \nabla_{(k}K_{ij)}=0\,, $$
	is called a \emph{Killing tensor field}, where round brackets denote symmetrization in enclosed indices,. By virtue of the metric $g$, these define examples of operator fields, given by 
	$$ K^i_j=g^{ia}K_{aj}\,, $$
	where we use the Einstein summation convention.
	By an abuse of notation we silently raise and lower indices, without introducing new symbols, as there is no risk of confusion.
	
	For an operator field $A$, its \emph{Nijenhuis tensor} (or \emph{Nijenhuis torsion}) is the $(1,2)$-tensor field defined by 
	\begin{equation}\label{eq:Nijenhuis}
		N_{jk}^i = \nabla_aA^i_kA^a_j-\nabla_aA^i_jA^a_k+(\nabla_kA^a_j-\nabla_jA^a_k)A^i_a\,.
	\end{equation}
	The \emph{Haantjes tensor} (or \emph{Haantjes torsion}) of $A$ then is the $(1,2)$-tensor field defined by
	\begin{equation}\label{eq:Haantjes}
		H_{jk}^i = N^b_{jk}A^i_aA^a_b+N^i_{ab}A^a_jA^b_k-A^i_a(N^a_{bk}A^b_j+N^a_{jb}A^b_k)\,.
	\end{equation}
	The Nijenhuis tensor gives rise to the Frölicher-Nijenhuis bracket, which is important in various contexts, such as separable systems and almost-complex structures \cite{NN,KMS}.
	The Fr\"olicher-Nijenjuis bracket is the first of an infinite class of brackets, the second of which is linked to Haantjes tensors \cite{TT}. The vanishing of the Haantjes tensor is connected, for instance, to the integrability of eigen-distributions of $A$, the integrability of hydrodynamic chains, the existence of separation coordinates, and the integrability of systems of partial differential equations (PDEs) of hydrodynamic type \cite{Haantjes,KKM,PSS,FM}.
	
	For an operator field $A$, a non-vanishing, closed differential form $\theta$ is said to be a \emph{conservation law}, if
	\begin{equation}
		d(A^{\ast}\theta) = 0    
	\end{equation}
	holds, i.e.\ if the $1$-form $A^{\ast}\theta$ is closed \cite{BKM_NijGeoIV}. Conservation laws play an important role in the theory of PDEs of hydrodynamic type \cite{KKM,Tsarev}.
	Conservation laws in $n$-component first-order PDEs with third-order
	Hamiltonian structures were recently discussed in \cite{FPV}.
	
	In the present paper, we are going to consider the following key question concerning the link between Haantjes-zero operator fields and the existence of conservation laws.
	\begin{question}\label{q:main.question}
		Is the Haantjes tensor for an operator field on an $n$-dimensional manifold zero, if it admits $n+1$ linearly independent conservation laws?
	\end{question}

	\section{Operator fields with \texorpdfstring{$n+1$}{n+1} conservation laws}
	
	Let us consider $(M,g)$ with $g=\sum_{k=1}^ndx_k^2$.
	Let $A$ be an operator field of the form
	$$ A=\sum_{j,k=1}^n a_{jk} \partial_{x_j}\otimes dx_k$$
	where $a_{jk}$, $1\leq j,k\leq n$, are functions on $M$.
	We consider the conservation laws generated by the $n+1$ functions
	\begin{align*}
		u^{(0)} &= \sum_{k=1}^n x_k^2\,,
		& \text{and} &&
		u^{(m)} &= x_m\,,\qquad\text{for $1\leq m\leq n$}.
	\end{align*}
	The condition that $A$ admits the conservation laws $du^{(m)}$, $0\leq m\leq n$, implies that $dA^{\ast}du^{(m)}=0$. This leads to the conditions
	\begin{subequations}
		\begin{align}
			0 &=
			\sum_{b=1}^n\left( a_{bj,k}u^{(0)}_{,b}-a_{bk,j}u^{(0)}_{,b}+a_{bj}u^{(0)}_{,bk}-a_{bk}u^{(0)}_{,bj} \right)
			\label{eq:hessian.first}
			\\
			0 &= a_{mj,k}-a_{mk,j}
			\label{eq:hessian.second}
		\end{align}
	\end{subequations}
	for $1\leq m,j,k\leq n$, where a comma denotes covariant differentiation.
	The conditions~\eqref{eq:hessian.second} yield
	\begin{equation}\label{eq:a.codazzi-solution}
		a_{ij}=a_{i,j}
	\end{equation}
	which we re-substitute into the condition~\eqref{eq:hessian.first}, implying
	\[
	0 = \sum_{b=1}^n\left[
	\left( a_{b,jk}-a_{b,kj}\right)x_{b}
	\right]+a_{k,j}-a_{j,k}
	= a_{k,j}-a_{j,k}
	\]
	where we use~\eqref{eq:a.codazzi-solution} and the Ricci identity in the second step.
	We conclude that
	\begin{equation}\label{eq:hessian.a}
		a_{ij}=f_{,ij}
	\end{equation}
	for a function $f$ on $M$.
	We are now asking which of these operator fields satisfying~\eqref{eq:hessian.a} have vanishing Haantjes tensor. 
	Using~\eqref{eq:Nijenhuis} and~\eqref{eq:Haantjes}, we have
	\[
	N_{jk}^i = \sum_a \left( f_{,ika}f_{,aj}-f_{,aij}f_{,ak}+(f_{,ajk}-f_{,akj})f_{,ia} \right)
	= \sum_a \left( f_{,ika}f_{,aj}-f_{,aij}f_{,ak} \right)
	\]
	and then
	\begin{equation}
		H_{jk}^i = \sum_{a,b} \left( N^b_{jk}f_{,ia}f_{,ab}+N^i_{ab}f_{,aj}f_{,bk}-f_{,ia}(N^a_{bk}f_{,bj}+N^a_{jb}f_{bk}) \right)\,.
	\end{equation}
	Our purpose here is not to solve the condition $H^i_{jk}=0$ in all generality. Instead, we shall give two examples demonstrating that the Haantjes tensor of an operator field admitting $n+1$ conservation laws can be vanishing or non-vanishing.
	To begin with, it is not hard to find a function $f$ such that all components $H^i_{jk}$ of the Haantjes tensor vanish. Indeed, in dimension two, an operator field of the form $A$ always has a vanishing Haantjes tensor. 
	\begin{example}
		The operator $A=\sum_{i,j=1}^3 \frac{\partial^2 f}{\partial x_i\partial x_j}\,\partial_{x_i}\otimes dx_j$ where
		\[
		f = x_1^3
		\]
		admits the $n+1=4$ conservation laws generated by $u^{(k)}$, with
		\[
		u^{(0)}=x_1^2+x_2^2+x_3^2\,,\qquad u^{(j)}=x_j\quad 1\leq j\leq 3\,,
		\]
		and has vanishing Haantjes tensor.
	\end{example}
	Next, we present a solution with non-vanishing Haantjes tensor.
	\begin{example}
		The operator $A=\sum_{i,j=1}^3 \frac{\partial^2 f}{\partial x_i\partial x_j}\,\partial_{x_i}\otimes dx_j$ where
		\[
		f = x_1^3+x_1x_2x_3
		\]
		admits the $n+1=4$ conservation laws generated by $u^{(k)}$, with
		\[
		u^{(0)}=x_1^2+x_2^2+x_3^2\,,\qquad u^{(j)}=x_j\quad 1\leq j\leq 3\,,
		\]
		but its Haantjes tensor is non-zero. Its non-vanishing components are
		\[
		H^b_{32}=-H^b_{23}=2x_1(x_1^2-x_2^2)
		\]
		for $b\in\{1,2,3\}$.
	\end{example}
	We can therefore conclude, regarding Question~\ref{q:main.question}, that the answer is generally negative, i.e.~the existence of $n+1$ conservation laws does not imply that an operator field is Haantjes-zero.
	Given the generality of Question~\ref{q:main.question}, this finding is, of course, not surprising. We will therefore now sharpen the hypothesis, asking the analogous question in the more restricted setting of operator fields arising from Killing tensor fields.

	\section{Killing tensor fields}\label{sec:SWI}
	
	We consider the restricted setting, in which the operator fields are required to be Killing tensor fields, identifying $(1,1)$-tensor fields and $(0,2)$-tensor fields by virtue of the metric.
	We ask the analog of Question~\ref{q:main.question}:
	\begin{question}\label{q:question.killing}
		Is the Haantjes tensor for a Killing tensor field zero if it admits $n+1$ linearly independent conservation laws?
	\end{question}
	
	In order to answer this question, consider a 3-dimensional flat manifold $(M,g)$ with $g=dx_1^2+dx_2^2+dx_3^2$, and the family of all second order Killing tensors on $M$ that admit four conservation laws generated by the functions 
	\begin{equation}\label{eq:potential.SWI}
		\begin{gathered}
			u^{(0)} = x_1^2+x_2^2+x_3^2\,,
			\\
			u^{(1)} = \frac{1}{x_1^2}\,,
			\qquad
			u^{(2)} = \frac{1}{x_2^2}\,,
			\qquad
			u^{(3)} = \frac{1}{x_3^2}\,.
		\end{gathered}
	\end{equation}
	We remark that, in dimension two, any Killing tensor field has vanishing Haantjes tensor.
	Since we work on flat (Euclidean) space, Killing tensor fields are reducible and arise via symmetric products of Killing vector fields.
	The space of Killing tensor in dimension three thus is $20$-dimensional \cite{T86} and an arbitrary Killing tensor field can therefore be parametrized in the form
	$$K:=\sum_{i=1}^{20} b_iK^{(i)}$$
	where $b_i\in\mathbb{R}$ for all $i=1,\dots,20$ and where $K^{(1)},\dots,K^{(20)}$ denote the symmetric product of the Killing vector fields, identified with their corresponding $1$-forms by virtue of the metric $g$.
	Since the $u^{(r)}$, $0\leq r\leq 3$, are required to yield conservation laws for $K$, we impose the condition
	\begin{equation}\label{eq:dKdV}
		d(K^{\ast}(du^{(r)}))=0
	\end{equation}
	where $K$ is the endomorphism identified with the Killing tensor field $K$ and where the $u^{(r)}$ are given in \eqref{eq:potential.SWI}. 
	We are hence left with a $6$-dimensional linear space of Killing tensor fields, generated by
	\begin{equation}
		\left.\begin{array}{lcl}
			K^{(1)}:=dx_1^2\,, && K^{(4)}:=(x_1dx_2-x_2dx_1)^2\,,
			\\
			K^{(2)}:=dx_2^2\,, && K^{(5)}:=(x_1dx_3-x_3dx_1)^2\,,
			\\
			K^{(3)}:=dx_3^2\,, && K^{(6)}:=(x_2dx_3-x_3dx_2)^2\,,
		\end{array}\right\}
	\end{equation}
	after reordering the indices of the $K^{(i)}$. The local form of an element in this space, in the coordinates $(x_1,x_2,x_3)$, is
	\begin{equation}\label{eq:K.SWI}
		K=\left(\begin{array}{rrr}
			b_{4} x_{2}^{2} + b_{5} x_{3}^{2} + b_{1} & -b_{4} x_{1} x_{2} & -b_{5} x_{1} x_{3} \\
			-b_{4} x_{1} x_{2} & b_{4} x_{1}^{2} + b_{6} x_{3}^{2} + b_{2} & -b_{6} x_{2} x_{3} \\
			-b_{5} x_{1} x_{3} & -b_{6} x_{2} x_{3} & b_{5} x_{1}^{2} + b_{6} x_{2}^{2} + b_{3}
		\end{array}\right).
	\end{equation}
	It is easily confirmed that the Haantjes tensor of such a $K$ is generically non-zero, as also the following example illustrates.
	\begin{example}
		Setting 
		$$b_2=b_1=0, \quad b_3=1, \quad b_4=b_5=2, \quad b_6=4,$$
		we obtain the operator field
		\begin{equation}\label{eq:K.not.haantjes.zero}
			\left(\begin{array}{rrr}
				x_{2}^{2} + x_{3}^{2} & -x_{1} x_{2} & -x_{1} x_{3} \\
				-x_{1} x_{2} & x_{1}^{2} + 2 \, x_{3}^{2} & -2 \, x_{2} x_{3} \\
				-x_{1} x_{3} & -2 \, x_{2} x_{3} & x_{1}^{2} + 2 \, x_{2}^{2} + 1
			\end{array}\right)\,.
		\end{equation}
		Its Haantjes tensor does not vanish and specifically $H^i_{jk}\ne0$ for $j\ne k$.
	\end{example}
	
	\begin{remark}\label{rmk:6b.SW1}
		We remark that the operator field~\eqref{eq:K.not.haantjes.zero} is non-degenerate, but does not satisfy the condition (6b) in~\cite{FPV}, i.e. 
		$$
		K_{m[k,n]l} = -\frac13 K^{pq}K_{p[l,m]}K_{q[k,n]}, 
		$$
		implying that it is not connected with a system of conservation laws with third-order Hamiltonian structures.
		
		Generally, the family~\eqref{eq:K.SWI} of operator fields contains non-degenerate as well as degenerate operator fields. We find that the Haantjes tensor for the degenerate realizations of~\eqref{eq:K.SWI} always vanishes.
		The non-degenerate realizations of~\eqref{eq:K.SWI} can have vanishing and non-vanishing Haantjes tensor. However, if a non-degenerate operator field~\eqref{eq:K.SWI} falls under condition~(6b) of~\cite{FPV}, then it has the form $b_1dx_1^2+b_2dx_2^2+b_3dx_3^2$ and hence its Nijenhuis tensor vanishes, implying that it is Haantjes-zero.
	\end{remark}
	
	For $K$ as in~\eqref{eq:K.SWI}, the Haantjes tensor vanishes if the parameters satisfy certain algebraic conditions, leading to the polynomial ideal $I\subset \mathbb Q[b_1,b_2,b_3,b_4,b_5,b_6]$,
	$$
	I = \langle~b_4\,J, (b_5+b_6)\,J,(b_3-b_2)\,J,(b_1-b_2)\,J,b_6\,J~\rangle\,,
	$$
	where
	$$
	J = b_2b_4b_5-b_3b_4b_5-b_1b_4b_6+b_3b_4b_6+b_1b_5b_6-b_2b_5b_6.
	$$
	The radical ideal of $I$ is
	$$ I_\text{rad} = \langle J\rangle\,,$$
	which is a primary ideal of Hilbert dimension~$5$.
	
	We hence conclude that the answer to Question~\ref{q:question.killing} is negative.
	We therefore aim to tighten our assumptions further. To this end, note that the conservation laws $u^{(r)}$ can be interpreted as potentials and that~\eqref{eq:dKdV} is the Bertrand-Darboux condition \cite{darboux_1901, bertrand_1857} for integrals in a second-order superintegrable system.
	Since any member of the family $K$ is a Killing tensor field~\eqref{eq:K.SWI}, these conditions ensure that the family
	\begin{equation}\label{eq:family.F.SWI}
		F = \sum_{i,j=1}^n K^{ij}p_ip_j+W
	\end{equation}
	where
	$$ dW = K^{\ast}d\left( \sum_{r=0}^n c_ru^{(r)} \right) $$
	is a family of integrals of motion for the Hamiltonian
	$$ H = \sum_{k=1}^n p_k^2+\sum_{r=0}^n c_ru^{(r)} $$
	where $c_r$ are real constants.
	Hence the canonical Poisson bracket of $H$ and $F$ vanishes,
	\[ \{H,F\}=0\,. \]

	\subsection{The Smorodinski-Winternitz I system}
	
	The family $F$ of integrals of the motion, c.f.~\eqref{eq:family.F.SWI}, is known to be the one associated to the second-order maximally superintegrable system usually referred to as the \emph{Smorodinski-Winternitz I} system \cite{KKM-III,Capel}.
	We therefore say that $K$ is the family of Killing tensors associated to the Smorodinski-Winternitz I system.
	
	Our discussion above therefore yields the following: Firstly, not every Killing tensor field associated to the Smorodinski-Winternitz I system has vanishing Haantjes tensor. Secondly, the subset of Killing tensor fields with vanishing Haantjes tensor within the (linear) space of all Killing tensor fields associated to the Smorodinski-Winternitz I system in dimension three forms an algebraic variety defined by a primary, radical ideal of Hilbert dimension five. This ideal does not contain any $5$-dimensional linear subspace of Killing tensors with vanishing Haantjes tensor. Indeed, by a direct investigation of the ideal $I_\text{rad}$, we obtain the following branches of solutions:\label{p:SWI.solution}
	\begin{enumerate}[label=(\roman*)]
		\item The inequalities $b_5\ne b_4$ and $b_6\ne0$ hold
		and $$b_1 = \frac{ (b_2-b_3) b_4 b_5 + (b_3 b_4 - b_2 b_5)b_6 }{ (b_4-b_5)b_6 }$$
		\item The equations $b_6=0$ and $b_4\ne b_5$ hold, and specifically one of the following cases is realized:
		\begin{itemize}
			\item The equations $b_6=0$, $b_3=b_2$ and the inequality $b_5\ne b_4$ hold
			\item The equations $b_6=b_4=0$ and the inequality $b_5\ne b_4$ hold.
			\item The equations $b_6=b_5=0$ and the inequality $b_5\ne b_4$ hold.
		\end{itemize}
		\item The equations $b_6=0$ and $b_5=b_4$ hold, together with
		$(b_3-b_2)b_4^2 = 0$. Specifically, one of the following cases is realized:
		\begin{itemize}
			\item The equations $b_4=b_5=b_6=0$ hold.
			\item The equations $b_2=b_3$, $b_6=0$, $b_4=b_5$ hold.
		\end{itemize}
		\item The equations $b_6\ne0$ and $b_4=b_5$ hold, i.e.\
		$0 = b_4(b_2-b_3)(b_4-b_6)$, and specifically, one of the following cases is realized:
		\begin{itemize}
			\item The equations $b_4=b_5=b_6$ and the inequality $b_6\ne0$ hold.
			\item The equations $b_2=b_3$, $b_4=b_5$ and the inequality $b_6\ne0$ hold.
			\item The equations $b_4=b_5=0$ and the inequality $b_6\ne0$ hold.
		\end{itemize}
	\end{enumerate}
	We therefore conclude that there are no 5-dimensional linear subspaces within the variety defined by the ideal $I$. We are going to pursue this line of reasoning further in the next section.

	\subsection{Second-order super\-inte\-gra\-ble systems with Haantjes-zero Killing tensor fields}
	
	We have found that there exist non-maximal second-order superintegrable systems with Haantjes-zero Killing tensor fields. 
	Let $(x_1,x_2,x_3,p_1,p_2,p_3)$ be the canonical Darboux coordinates on $T^{\ast}M$. The integrals of motion
	\begin{align*}
		F^{(1)}&=p_1^2+\frac{a_1}{x_1^2}+x_1^2a_0 \\
		F^{(2)}&=p_2^2+\frac{a_2}{x_2^2}+x_2^2a_0 \\
		F^{(3)}&=p_3^2+\frac{a_3}{x_3^2}+x_3^2a_0 \\
		F^{(5)}&=(x_3p_1-x_1p_3)^2 + x_3^2\frac{a_1}{x_1^2} + x_1^2\frac{a_3}{x_3^2}
	\end{align*}
	(compare the coefficients in~\eqref{eq:K.SWI} of $b_1$ to $b_3$ and of $b_5$) are functionally independent and are, in the sense of Equation~\eqref{eq:dKdV}, compatible with the potential
	$$ V=a_0\,(x_1^2+x_2^2+x_3^2)+\frac{a_1}{x_1^2}+\frac{a_2}{x_2^2}+\frac{a_3}{x_3^2}+a_4. $$
	This potential, together with the integrals $F^{(1)},F^{(2)},F^{(3)}$ and $F^{(5)}$ provide an example of a (non-maximal) second-order superintegrable system in dimension three. Since these systems can be viewed as restrictions of the Smorodinski-Winternitz I system, the discussion in Section~\ref{sec:SWI} shows that, for any choice of the parameters, any Killing tensor field of the form
	$$ K_{b_1,b_2,b_3,b_5}=b_1dx_1^2+b_2dx_2^2+b_3dx_3^2+b_5(x_3dx_1-x_1dx_3)^2$$
	has vanishing Haantjes tensor.
	
	We constructed the example above using the equations $b_6=b_4=0$, a case of the solutions $(ii)$ written above in Section~\ref{sec:SWI}. One can find more examples of this kind, analyzing the other conditions listed above, see page~\pageref{p:SWI.solution}, confirming the existence of (non-maximal) second-order superintegrable system whose associated Killing tensor fields are all Haantjes-zero.
	In the next section, we will therefore sharpen our restrictions and focus the investigation on the case of second-order \emph{maximally} superintegrable systems.

	\section{Killing tensor fields in non-degenerate maximally superintegrable systems of second order}

	Non-degenerate second-order maximally superintegrable systems in dimension three (and two) are classified \cite{KKM-III,KKM-IV,Capel,Evans,KKPM}, and we can therefore, in dimension three, extend the line of reasoning started in the previous section, where we found that the Killing tensor fields associated to the Smorodinski-Winternitz system I can have vanishing or non-vanishing Haantjes tensor.
	Moreover, the considerations in the previous section have shown that there exist \emph{non-maximal} superintegrable systems of second order such that the associated integrals of the motion arise from Haantjes-zero Killing tensors.
	Despite this fact, however, there does not exist a $5$-dimensional linear subspace of Killing tensors compatible with the potential of the Smorodinski-Winternitz I system and such that the Haantjes tensor for these Killing tensors vanishes.
	
	
	In the present section, we are going to investigate the question analogous to Question~\ref{q:main.question} for all second-order maximally superintegrable systems:
	\begin{question}\label{q:question.maximal.sis}
		Do second-order maximally superintegrable systems exist such that all associated Killing tensor fields have vanishing Haantjes tensor?
	\end{question}

	\subsection{Three-dimensional non-degenerate systems}
	
	We will find that the answer to this question is positive in dimension three, but that all such systems are projections of the so-called \emph{abundant} isotropic harmonic oscillator system.
	
	From the classification for dimension three, it follows that there exist, on Euclidean $3$-dimensional spaces, four distinct families of superintegrable potentials: the isotropic harmonic oscillator potential, the Smorodinski-Winternitz I potential, and the two ``mixed'' Smorodinski-Winternitz II potentials, labelled by OO and IV in \cite{Capel}.
	In the current section, we dedicate ourselves to investigating linear subspaces of Haantjes-zero Killing tensor fields compatible with the mentioned superintegrable potentials via~\eqref{eq:dKdV}. We ask whether we can find these subspaces such that a basis of Killing tensors exists whose associated integrals of the motion (including the Hamiltonian) are functionally independent.
	
	\subsubsection{Smorodinski-Winternitz I}
	
	We have already studied the Smorodinski-Winternitz I potential in Section~\ref{sec:SWI}, and have found that there is no $5$-dimensional linear subspace within the space of compatible Killing tensor fields of the potential that satisfies the desired properties.
	
	\subsubsection{Isotropic Harmonic Oscillator}
	
	We therefore continue our discussion with the $(n+1)$-parameter isotropic harmonic oscillator potential, i.e.\ system O in \cite{Capel}, 
	$$ V_O=a_0\,(x_1^2+x_2^2+x_3^2)+a_1\,x_1+a_2\,x_2+a_3\,x_3+a_4 $$
	where we ignore the constant term in the potential as it does not yield a non-trivial conservation law.
	
	The space of compatible Killing tensors of the potential is parametrized by
	\begin{align*}
		K = \begin{pmatrix}
			b_1 & b_4& b_5 \\
			b_4 & b_2 & b_6 \\
			b_5 & b_6 & b_3
		\end{pmatrix}.
\end{align*}
A straightforward computation shows that the Haantjes tensor is zero for any member in this family of Killing tensors (identified with the associated operator fields by virtue of the metric $g$).

In fact, as we will prove further below, the analogous statements holds true for the isotropic harmonic oscillator potential in any dimension.

In the next two paragraphs, we will confirm that the harmonic oscillator potential, in dimension three, is the only system with this property, and hence that the following claim holds.
We shall now classify all the non-degenerate maximally superintegrable systems, defined by five functionally independent integrals of motion $F^{(r)}=K^{(r)ij}p_ip_j+W^{(r)}$, $0\leq r\leq 4$ compatible with a 5-dimensional space of potentials, such that the Killing tensors $\sum_rc_{(r)}K^{(r)}_{ij}$ have vanishing Haantjes tensor.

\begin{theorem}
	A non-degenerate second-order maximally superintegrable system in dimension three, defined by a $5$-dimensional space of Killing tensors $\mathcal K$, whose Killing tensors are all Haantjes-zero, is compatible with the isotropic harmonic oscillator potential
	$$ V=a_0(x^2+y^2+z^2)+a_1x+a_2y+a_3z+a_4. $$
	and is defined by a subspace $\mathcal K'\subset\mathcal K$ of dimension five that contains the metric~$g$ and admits a basis of Killing tensors such that the associated integrals of motion (including the Hamiltonian) are functionally independent.
\end{theorem}

\subsubsection{Smorodinski-Winternitz II, first example}

We begin with the system OO of~\cite{Capel},
$$ V_{OO}=a_0\,(4x_1^2+4x_2^2+x_3^2)+a_1\,x_1+a_2\,x_2+\frac{a_3}{x_3^2}+a_4. $$
A compatible Killing tensor is of the local form
$$ K = \begin{pmatrix}
	b_1 & b_5 & -b_4 x_3 \\
	b_5 & b_2 & -b_6 x_3 \\
	-b_4x_3 & -b_6 x_3 & 2b_4x_1+2b_6x_2+b_3
\end{pmatrix}\,. $$
As we did in Section~\ref{sec:SWI}, for the Smorodinski-Winternitz I system, we can analogously compute the polynomial ideal $I\subset \mathbb Q[b_1,b_2,b_3,b_4,b_5,b_6]$ describing the Haantjes-zero Killing tensor fields within the family $K$ now under consideration.
The associated radical ideal is
$$ I_\text{rad} = \langle\quad b_1b_4b_6-b_2b_4b_6-b_4^2b_5+b_5b_6^2\quad\rangle $$
and describes the variety of Haantjes-zero Killing tensor fields, which is naturally embedded into the space of all Killing tensor fields of $g$. 
We find that $I_\text{rad}$ is a primary ideal and that its Hilbert dimension is $\dim(I_\text{rad})=\dim(I)=5$. However, the variety defined by $I$ is not a linear subspace, nor can it contain any linear subspace of dimension five.

To conclude, we remark that, unlike the case of the Smorodinski-Winternitz~I system (see Remark~\ref{rmk:6b.SW1}), the Haantjes tensor is not necessarily zero, even if (6b) from~\cite{FPV} holds.

\subsubsection{Smorodinski-Winternitz II, second example}

It remains to consider the system IV of~\cite{Capel},
$$ V_{IV}=a_0\,(4x_1^2+x_2^2+x_3^2)+a_1\,x_1+\frac{a_2}{x_2^2}+\frac{a_3}{x_3^2}+a_4. $$
A compatible Killing tensor is of the local form
$$ K = \begin{pmatrix}
	b_1 & -b_6x_2 & -b_4x_3 \\
	-b_6x_2 & b_5x_3^2 + 2b_6x_1 + b_2 & -b_5x_2x_3 \\
	-b_4x_3 & -b_5x_2x_3 & b_5x_2^2+2b_4x_1+b_3
\end{pmatrix}\,. $$
As before, we compute the polynomial ideal $I\subset \mathbb Q[b_1,b_2,b_3,b_4,b_5,b_6]$ describing the Haantjes-zero Killing tensor fields within the family $K$.
The associated radical ideal is
$$ I_\text{rad} = \langle\quad b_1b_4b_5 - b_2b_4b_5 + b_4^2b_6 - b_1b_5b_6 + b_3b_5b_6 - b_4b_6^2 \quad\rangle, $$
which is a primary ideal.
Its Hilbert dimension is $\dim(I_\text{rad})=\dim(I)=5$.
The variety defined by $I$ is not a linear subspace, nor can it contain any linear subspace of dimension five.

To conclude, we remark that, as in the case of the Smorodinski-Winternitz I system (see Remark~\ref{rmk:6b.SW1}), the Haantjes tensor is necessarily zero, if (6b) from~\cite{FPV} holds.
The answer to Question~\ref{q:question.maximal.sis} is therefore negative.

\subsection{Abundant systems in any dimension}
All \emph{$3$-dimensional} non-degenerate second-order maximally superintegrable systems are abundant \cite{KKM-III,M}, and it is therefore instructive to investigate the class of abundant second-order superintegrable systems further.
In the present section, we are going to find that the Killing tensor fields compatible with the potentials of an isotropic harmonic oscillator potential are always Haantjes-zero, while the Killing tensors of other systems might or might not have vanishing Haantjes tensor.
The non-degenerate isotropic harmonic oscillator system therefore is the only family of arbitrary-dimensional second-order maximally superintegrable systems that has this property.

The discussion in the current paragraph relies on the geometric framework developed for irreducible second-order maximally superintegrable systems in~\cite{KSV23,KSV24}.
In particular, in the reference it is shown that for an abundant second-order maximally superintegrable system in dimension $n\geq3$, the associated Killing tensor fields satisfy
\[
\nabla_kK_{ij} = P_{ijk}^{ab}K_{ab}
\]
for a unique tensor field $P^{ab}_{ijk}$, called the \emph{structural tensor} of the system.
The tensor field $P^{ab}_{ijk}$ is going to be crucial for the following discussion.
For a concise notation, we introduce the auxiliary tensor field
\[
Q^{ai}_{bjk}=g^{ic}g_{bd}P^{ad}_{cjk}
\]
and hence obtain, for the Nijenhuis tensor of $K$,
\begin{align*}
	N_{jk}^i
	&= \nabla_aK^i_kK^a_j-\nabla_aK^i_jK^a_k+(\nabla_kK^a_j-\nabla_jK^a_k)K^i_a \\
	&= Q^{ic}_{dak}K^d_cK^a_j-Q^{ci}_{daj}K^d_cK^a_k+Q^{ca}_{dkj}K^d_cK^i_a-Q^{ca}_{djk}K^d_cK^i_a \\
	&= \left( Q^{ic}_{dak}g^b_j-Q^{ci}_{daj}g^b_k
	+Q^{cb}_{dkj}g^i_a-Q^{cb}_{djk}g^i_a \right) K^d_cK^a_b\,.
\end{align*}
After introducing the abbreviation
\[
U^{icb}_{dakj} := Q^{ic}_{dak}g^b_j-Q^{ci}_{daj}g^b_k
+Q^{cb}_{dkj}g^i_a-Q^{cb}_{djk}g^i_a\,,
\]
we may write this formula more concisely as
\[
N^i_{jk} = U^{icb}_{dakj} K^d_cK^a_b
\]
It then follows for the Haantjes tensor of $K$ that
\begin{align*}
	H_{jk}^i
	&= N^b_{jk}K^i_aK^a_b
	+N^i_{ab}K^a_jK^b_k
	-K^i_aN^a_{bk}K^b_j
	-K^i_aN^a_{jb}K^b_k \\
	&= U^{rcb}_{dakj} K^d_cK^a_bK^m_rK^i_m
	+U^{icm}_{drab}K^d_cK^m_rK^a_jK^b_k
	\\ &\qquad
	-U^{acm}_{drbk}K^d_cK^r_mK^i_aK^b_j
	+U^{acm}_{drjb}K^d_cK^r_mK^i_aK^b_k \\
	&= \bigg(
	U^{mcb}_{dakj} g^a_pg^q_b g^i_\mu g^\nu_r 
	+U^{icm}_{drab} g^a_pg_j^q g^b_\mu g^\nu_k
	\\ &\qquad
	-U^{acm}_{drbk} g^i_pg^q_a g^b_\mu g^\nu_j
	+U^{acm}_{drjb} g^i_pg^q_a g^b_\mu g^\nu_k
	\bigg) K^d_cK^r_mK^p_q K^\mu_\nu
\end{align*}
and therefore, introducing the abbreviation
\begin{align*}
	\mathcal U^{i\ cmq\nu}_{jk\ drp\mu} &=
	U^{mcb}_{dakj} g^a_pg^q_b g^i_\mu g^\nu_r 
	+U^{icm}_{drab} g^a_pg_j^q g^b_\mu g^\nu_k
	\\ &\qquad
	-U^{acm}_{drbk} g^i_pg^q_a g^b_\mu g^\nu_j
	+U^{acm}_{drjb} g^i_pg^q_a g^b_\mu g^\nu_k\,,
\end{align*}
we arrive at
\begin{equation}\label{eq:Haantjes.abundant}
	H^i_{jk} = \mathcal U^{i\ cmq\nu}_{jk\ drp\mu} K^d_cK^r_mK^p_q K^\mu_\nu
\end{equation}
where $\mathcal U^{i\ cmq\nu}_{jk\ drp\mu}$ does not depend on the choice of $K$.\footnote{%
	To prevent misconceptions, we recall that the structural tensor \smash{$P^{ab}_{ijk}$}, and hence also $\mathcal U$, depends on the (whole) space $\mathcal K$ of all associated Killing tensor fields of the system, c.f.\ \cite{KSV23,KSV24}. The point here is that $\mathcal U$ is determined by the system in its entirety, while $K$ in~\eqref{eq:Haantjes.abundant} is a particular choice of a Killing tensor field from the space $\mathcal K$.}
Specifically, we observe that for the (abundant) isotropic harmonic oscillator system, $P_{ijk}^{ab}=0$ and hence $\nabla K=0$ for all Killing tensor fields $K$ associated to the system. It follows that the Haantjes tensor of $K$ vanishes, $H_K=0$, for any Killing tensor field $K$ associated to the (abundant) isotropic harmonic oscillator system, regardless of dimension. Concerning dimension, we recall that the vanishing of the Haantjes tensor is automatic in dimension two. The following example summarizes the discussion and provides an answer to Question~\ref{q:question.maximal.sis}.
\begin{example}
	If $K$ is a Killing tensor field compatible with the non-degenerate isotropic harmonic oscillator potential, then its Haantjes tensor is necessarily zero, regardless of dimension.
	
	The non-degenerate isotropic harmonic oscillator family of systems is the only arbitrary-dimensional family of second-order maximally superintegrable system for which this property holds in \emph{any} dimension.
\end{example}

The second statement in the example follows from the earlier discussion in the present section, where we have seen that in dimension three the isotropic harmonic oscillator potential is the only non-degenerate second-order maximally superintegrable potential whose associated Killing tensors all have vanishing Haantjes tensor. With Equation~\eqref{eq:Haantjes.abundant} at hand, this is indeed not surprising: note that $\mathcal U$ is linear in the structural tensor $P$, and hence $\mathcal U=0$ implies $P=0$. This, in turn, implies that the \emph{structure tensor} of the system, introduced in \cite{KSV23,KSV24}, vanishes, implying $\nabla^2V=\frac1n\,g\,\Delta V$ for the Hessian $\nabla^2$ and Laplacian $\Delta V$ of the superintegrable potential. The general solution of this system, on flat space, is precisely the non-degenerate isotropic harmonic oscillator potential. The isotropic harmonic oscillator therefore is the only example for which the Haantjes tensor vanishes in the completely generic manner as in the example. 

\section*{Acknowledgements}

We thank the participants and organisers of MATRIX-SMRI Symposium \emph{Nijenhuis Geometry and Integrable Systems II} and the Nijenhuis preworkshop at La Trobe University. In particular, we are grateful towards Vladimir Matveev, Mike Eastwood, Andrey Konyaev, Konrad Schöbel and Jonathan Kress and Yuri Nikolajevski for discussions and insights.

This work has been funded by the German Research Foundation through the project grants \#540196982, \#455806247. The research of Ian Marquette was supported by Australian Research Council Future Fellowship FT180100099.

\end{document}